\documentclass[%
 reprint,
 amsmath,amssymb,
 aps,
pra,
]{revtex4-2}

\usepackage{graphicx}% Include figure files
\usepackage{dcolumn}% Align table columns on decimal point
\usepackage{amsmath,amssymb,bm,booktabs}
\usepackage{soul}
\usepackage{xcolor}
\usepackage[english]{babel}
\usepackage{hyperref}
\begin{document}

% \preprint{APS/123-QED}

\title{Polarization-encoded photon statistics for counting dipole emitters}% Force line breaks with \\
% \thanks{A footnote to the article title}%

\author{Oumeng Zhang}
\email{ozhang@tamu.edu}
\affiliation{Department of Physics \& Astronomy, Institute for Quantum Science \& Engineering, Texas A\&M University.}

% \date{\today}% It is always \today, today,
             %  but any date may be explicitly specified

\begin{abstract}
Counting the emitters within a diffraction-limited region is a prerequisite for localizing overlapping molecules. Under pulsed excitation, each molecule emits at most one photon per pulse, so photon-number-resolved detection can count emitters without resolving them. But the required number of pulses grows steeply with the emitter number. Here we show that polarization, conventionally used to measure the orientations of dipole emitters, also carries information on their number. Because mean polarization-resolved signals are unchanged when every emitter is replaced by multiple dimmer copies, this information resides in the photon statistics within each pulse, which polarization reshapes in two ways. Linearly polarized excitation at three angles increases the sub-Poissonian deficit of the photon count by a factor of 1.5 for any emitter orientation, and detection in four polarization channels reveals the emitter number through how the photons of multiphoton pulses are distributed among the channels. Maximum-likelihood simulations for two to five emitters of unknown orientations and brightnesses show that three-angle excitation reduces the required number of pulses by about 2.3-fold, close to the predicted $1.5^2=2.25$, and that adding four-channel detection raises the reduction to as much as 4.8-fold. These gains require only polarization optics added to photon-number-resolved detection.
\end{abstract}

%\keywords{Suggested keywords}%Use showkeys class option if keyword
                              %display desired
\maketitle

%\tableofcontents

\section{Introduction}

Resolving emitters whose images overlap is a long-standing challenge in optical microscopy: fluorescent emitters closer than the diffraction limit ($\sim$250~nm) produce a single blurred spot. Single-molecule localization microscopy (SMLM) overcomes this through temporal sparsity: at most one emitter may be active within a diffraction-limited region in each frame, so that each emitter can be localized individually, with a precision of about 20~nm~\cite{Betzig2006,Rust2006,Hess2006,lelek_single-molecule_2021}. However, sparsity is also what makes SMLM slow. A reconstruction takes thousands of frames and minutes to hours, so nanoscale organization that changes faster, such as the transient membrane domains and receptor clusters that sort proteins and relay signals~\cite{sezgin_mystery_2017,groves_molecular_2010}, is recorded only as before-and-after snapshots~\cite{Lu2020,Zhang2022}. Allowing a moderate overlap, with a few emitters simultaneously active within a diffraction-limited region, would shorten the acquisition severalfold.
High-density localization algorithms can analyze partially overlapping images at densities of up to about 10~\textmu{}m$^{-2}$ \cite{holden_daostorm_2011,huang_simultaneous_2011,zhu_faster_2012}, about one active emitter per $(300~\mathrm{nm})^2$, comparable to the area of a diffraction-limited spot.
Within a diffraction-limited region, however, the image no longer reveals how many emitters it contains, and this number must be determined before the emitters can be localized.

Quantum estimation theory has revealed what can be resolved within such a region.
Tsang et al. showed that, with a suitable measurement, the separation of two incoherent point sources can be estimated with a precision that does not degrade as the sources approach each other~\cite{Tsang2016,tsang_resolving_2019}, a prediction since confirmed experimentally~\cite{paur_achieving_2016,tham_beating_2017}. This approach has recently been extended to pairs of dipole emitters~\cite{dingilian_quantifying_2026}.
These analyses rest on a critical premise: the number of sources is known, typically exactly two, and the authors of Ref.~\cite{dingilian_quantifying_2026} note that uncertainty in this number complicates the problem. In an experiment with overlapping molecules, the number is the first unknown: before positions or orientations can be estimated, one must decide whether a region contains two molecules, three, or more. Quantum hypothesis testing addresses this question only partly, having mainly been used to decide between one source and two~\cite{lu_quantum-optimal_2018,zanforlin_optical_2022}. 

Quantum optics offers a way to count emitters without resolving them. Because a single molecule cannot emit two photons at once, $N$ independent, equally bright single-photon emitters show antibunching, with $g^{(2)}(0)=1-1/N$~\cite{kimble_photon_1977,basche_photon_1992}.
Photon correlations have also been exploited in microscopy, to enhance resolution~\cite{schwartz_superresolution_2013,tenne_super-resolution_2019} and to localize emitters in non-sparse scenes by inferring how many are active~\cite{israel_quantum_2017}.
Estimating $g^{(2)}(0)$, however, relies on two-photon coincidences, which are rare because a fluorophore is detected in only a small fraction of excitation cycles, so such counting requires long acquisitions. Pulsed excitation with photon-number-resolving detection extracts more from the same light: maximum-likelihood counting uses the full distribution of the number of photons detected after each pulse, including pulses with none~\cite{li_en_2022,li_estimation_2023}. The number of pulses needed to tell $N$ from $N+1$ nevertheless grows steeply with $N$, because their statistics become increasingly alike. Whether this pulse budget can be reduced further is the question we address.

Fluorescent molecules are not point sources but dipoles, whose absorption and emission depend on their orientations. Modulating the excitation polarization reveals the in-plane orientation of a single molecule \cite{Ha1996}, and orientation-resolved SMLM now measures these orientations routinely \cite{brasselet_polarization_2023,zhang_single-molecule_2024,brasselet_single-molecule_2025}.
Polarization is therefore a natural candidate for separating or counting overlapping emitters. Excitation-polarization modulation was proposed to resolve overlapping emitters~\cite{hafi_fluorescence_2014}, but adds little information to mean-intensity super-resolution~\cite{frahm_polarization_2016}; photoselection makes the emission polarization of a small cluster depend on its number of molecules, which can reveal clustering in polarization histograms over many clusters~\cite{bell_transition_2015}; and combining excitation-polarization modulation with an engineered dipole-spread function can distinguish a pair of overlapping molecules from a single rotating one~\cite{chen_resolving_2025}. All these approaches rely on mean polarization signals.
As we show in Sec.~\ref{sec:theory}, such signals cannot count in general, even when resolved in both excitation and emission polarization: replacing every molecule by two copies with half its brightness leaves all of them unchanged.

Here we use polarization differently: not to measure orientations, but to reshape the within-pulse photon statistics from which the number of emitters is inferred. We consider pulsed excitation with photon-number-resolved readout and compare, at a fixed number of pulses, circular excitation with total photon counting against excitation with three linear polarizations, detection in four polarization-resolved channels, and their combination. 
Each polarization element has been used for single-molecule orientation measurements: excitation cycled among three linear polarizations~\cite{Backer2016}, emission split into four polarization channels~\cite{Rimoli2022}, and excitation-polarization modulation combined with polarization-resolved detection~\cite{Beausang2013}.
We show analytically that linearly polarized excitation enhances the counting contrast for any emitter orientations, and that polarization-resolved detection adds count information through the allocation of photons among channels in multiphoton pulses. Maximum-likelihood simulations for two to five emitters of unknown orientations and brightnesses confirm these predictions: excitation polarization substantially reduces the required pulse budget, whereas emission polarization helps mainly in combination with it, increasingly so as the number of emitters grows for equal amplitudes and tilted dipoles.

\section{Measurement protocols and forward model}
\label{sec:protocols}
We consider an unresolved cluster of emitters excited by laser pulses much shorter than the fluorescence lifetime, so that each emitter is excited at most once per pulse \cite{lounis_single_2000}. Fluorescence is collected in a detection window synchronized to each pulse. The pulse period greatly exceeds the lifetime; the emitters therefore return to the ground state between pulses, and every detected photon can be assigned to the pulse that produced it. The observable is the number of photons detected after each pulse, including pulses with no detected photon. Unlike a time-integrated intensity measurement, which reports only the mean photon number, this pulse-resolved record retains the within-pulse photon statistics. A few bright emitters and a larger number of dimmer ones can produce the same mean signal but different probabilities of multiphoton events, so these statistics can reveal the emitter number without a brightness calibration \cite{grusmayer_time-resolved_2017,li_en_2022}. We ask whether polarization-controlled excitation and polarization-resolved detection reduce the pulse budget needed to infer the emitter number when the individual emitter orientations and brightnesses are unknown.

\begin{figure}[t!]
\centering
\includegraphics[]{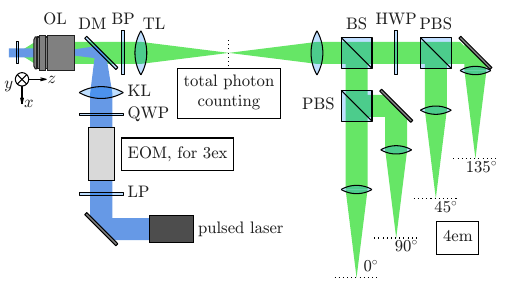}
\caption{Setup for polarization-encoded photon counting. Pulsed excitation (blue) passes a linear polarizer (LP), an electro-optic modulator (EOM), a quarter-wave plate (QWP), and a K\"ohler lens (KL), and a dichroic mirror (DM) reflects it into the objective lens (OL). For circular excitation, the EOM is inactive and the QWP is oriented at $45^\circ$ to the LP; for three-angle excitation (3ex), the QWP is parallel to the LP, and the EOM rotates the linear polarization to $\psi_p = 0^\circ$, $60^\circ$, and $120^\circ$. Fluorescence (green) passes a bandpass filter (BP) and a tube lens (TL). For total photon counting, the detector is placed at the intermediate image plane. For four-channel detection (4em), the recollimated fluorescence is divided by a non-polarizing 50:50 beam splitter (BS), and polarizing beam splitters (PBS) separate the $0^\circ$ and $90^\circ$ components in one arm and, after a half-wave plate (HWP), the $45^\circ$ and $135^\circ$ components in the other. Dotted lines mark the photon-number-resolved detection planes. Combining the two excitation and two detection schemes gives the baseline, 4em, 3ex, and 3ex+4em protocols. The optical axis is $z$, and all angles are measured from the $x$ axis in the sample plane.}
\label{fig:measurement}
\end{figure}

We compare four protocols (Fig.~\ref{fig:measurement}): circular excitation with total photon counting (baseline); circular excitation with four polarization-resolved emission channels (4em); three linear excitation polarizations with total photon counting (3ex); and their combination (3ex+4em). All four protocols use the same total number of pulses $M$. We label the excitation polarization by $p$, with $p=\mathrm{circ}$ for circular excitation. In 3ex and 3ex+4em, the pulses are divided equally among the three linear polarizations, $p=1,2,3$, with angles and pulse numbers
\begin{equation}
\psi_p=(p-1)\pi/3,\quad M_p=M/3.
\label{eq:settings}
\end{equation}

All four protocols require photon-number-resolved readout synchronized to the excitation pulses, with a separate readout for each emission channel in 4em and 3ex+4em. A scanning implementation could use a photon-number-resolving module built from multiplexed avalanche photodiodes \cite{achilles_fiber-assisted_2003}. For widefield imaging, photon-number-resolving cameras such as qCMOS sensors are an option if acquisition is synchronized so that each exposure contains fluorescence from a single pulse \cite{roberts_comparison_2024}. Throughout, we assume ideal photon-number-resolved readout and do not model a specific detector.

For every protocol, we represent the readout of one pulse at polarization $p$ by a count vector and its total,
\begin{equation}
\bm K_p=(K_{p1},\ldots,K_{pC})\in\mathbb N_0^C,\quad
K_{p,\mathrm{tot}}\equiv\sum_{c=1}^{C}K_{pc},
\label{eq:readout}
\end{equation}
where $K_{pc}$ is the number of photons detected in channel $c$. Total-only readout has a single channel ($C=1$), so $K_{p,\mathrm{tot}}=K_{p1}$; channel-resolved readout has $C=4$. Lowercase $\bm k$ and $k_{\mathrm{tot}}$ denote observed values of $\bm K_p$ and $K_{p,\mathrm{tot}}$.

We model the emitters as independent fixed dipoles with collinear absorption and emission moments and a linear (unsaturated) excitation response, neglecting energy transfer and annihilation between them. Background and blinking are not included.
For the $i$th emitter ($i=1,\ldots,N$), with its dipole in the sample ($xy$) plane at angle $\phi_i$ from the $x$ axis, the detection probability under linear excitation polarized at angle $\psi$ is
\begin{equation}
 q_i(\psi)=2a_i\cos^2(\phi_i-\psi)=a_i\{1+\cos[2(\phi_i-\psi)]\}.
 \label{eq:excitation}
\end{equation}
The amplitude $a_i$, which sets the brightness of emitter $i$, combines its absorption cross section and quantum yield with the excitation pulse energy and the collection and detection efficiencies; $2a_i$ is the detection probability for excitation polarized along the dipole. Circular excitation at the same pulse energy gives the angle average of Eq.~\eqref{eq:excitation}, $a_i$. Because the angles $2\psi_p$ are equally spaced around the circle, the three linear polarizations reproduce this average:
\begin{equation}
 \frac13\sum_{p=1}^{3}q_i(\psi_p)=a_i,\quad
 \bar\lambda\equiv\frac13\sum_{p=1}^{3}\mathbb{E}[K_{p,\mathrm{tot}}]=\sum_i a_i.
 \label{eq:meanbudget}
\end{equation}
All four protocols therefore have the same mean number of detected photons per pulse, $\bar\lambda$, and the same expected total, $M\bar\lambda$.

For channel-resolved readout, a lossless, non-polarizing 50:50 beam splitter divides the fluorescence between two arms, and polarizing beam splitters in the two arms separate the $0^\circ/90^\circ$ and $45^\circ/135^\circ$ components, respectively. Conditional on detection, a photon from the $i$th emitter enters channel $c$ with probability
\begin{equation}
 r_c(\phi_i)=\frac{1+V\cos[2(\phi_i-\gamma_c)]}{4},\quad \gamma_c=(c-1)\pi/4,
 \label{eq:routing}
\end{equation}
where $\gamma_c$ is the polarization angle selected by channel $c$, and the polarization visibility $V<1$ accounts for the polarization mixing caused by high-numerical-aperture collection.
Because $\sum_{c=1}^{4}r_c=1$, the split redistributes detected photons among the channels without changing their total.
For a numerical aperture of 1.4 and a homogeneous medium of refractive index $n=1.515$, high-aperture collection theory gives $V=0.958$ \cite{axelrod_carbocyanine_1979,Fourkas2001,Axelrod2012,Backer2014}.

In a pulse at polarization $p$, emitter $i$ contributes $\bm X_{ip}\in\mathcal X=\{\bm0,\bm e_1,\ldots,\bm e_C\}$ to $\bm K_p$, where $\bm e_c$ represents one photon detected in channel $c$. Let $q_{ip}$ be the detection probability of emitter $i$ at polarization $p$, with $q_{ip}=q_i(\psi_p)$ for $p=1,2,3$ and $q_{i,\mathrm{circ}}=a_i$. Let $\rho_{ic}$ be the probability that a detected photon from emitter $i$ enters channel $c$, with $\rho_{i1}=1$ for total-only readout and $\rho_{ic}=r_c(\phi_i)$ for channel-resolved readout. The single-emitter probability mass function (PMF) is then
\begin{equation}
 \Pr(\bm X_{ip}=\bm x)=
 \begin{cases}
 1-q_{ip}, & \bm x=\bm0,\\[2pt]
 q_{ip}\,\rho_{ic}, & \bm x=\bm e_c,\quad c=1,\ldots,C,
 \end{cases}
 \label{eq:singlepmf}
\end{equation}
where $\bm x=\bm0$ covers all cases in which no photon from emitter $i$ is detected.

The count vector is the sum of these contributions, $\bm K_p=\sum_i\bm X_{ip}$, and independence between emitters gives its PMF as a finite convolution,
\begin{equation}
 P_p(\bm k)\equiv\Pr(\bm K_p=\bm k)=
 \sum_{\substack{\bm x_1,\ldots,\bm x_N\in\mathcal X\\
                  \bm x_1+\cdots+\bm x_N=\bm k}}
 \prod_{i=1}^{N}\Pr(\bm X_{ip}=\bm x_i),
 \label{eq:jointpmf}
\end{equation}
where the sum runs over all combinations of single-emitter outcomes that add up to $\bm k$. For example, with two emitters and channel-resolved readout, the probability of exactly one photon, detected in channel 1, is
\begin{equation}
 P_p(1,0,0,0)=q_{1p}r_1(\phi_1)(1-q_{2p})+(1-q_{1p})q_{2p}r_1(\phi_2),
 \label{eq:pmfexample}
\end{equation}
with one term for each possible source emitter. All PMFs are conditional on $N$ and the emitter parameters $\{a_i,\phi_i\}$, which we omit from the notation for readability.

For total-only readout, Eq.~\eqref{eq:jointpmf} reduces to a Poisson binomial distribution with success probabilities $q_{ip}$, which becomes $\mathrm{Binomial}(N,a)$ for equal amplitudes $a_i=a$ under circular excitation. Since no emitter contributes more than one photon per pulse, $K_{p,\mathrm{tot}}\le N$ for either readout. For channel-resolved readout, the four channel counts within a pulse are negatively correlated, because each emitter's photon enters only one channel; Eq.~\eqref{eq:jointpmf} gives their joint distribution. Distinct pulses are independent given the emitter parameters, so the probability of a complete photon record is the product of single-pulse PMFs, each evaluated at the polarization of that pulse.

\section{How polarization adds counting information}
\label{sec:theory}
Throughout this section, we take all emitters to have a common amplitude $a$, for which the counting identities below are exact and show explicitly how polarization adds counting information; unequal amplitudes and three-dimensional orientations are treated with the full likelihood in Sec.~\ref{sec:inference}. We first consider the baseline protocol. Under circular excitation, the total count per pulse is binomial, $K_{\mathrm{circ},\mathrm{tot}}\sim\mathrm{Binomial}(N,a)$, with mean $\lambda_{\mathrm{circ}}=Na$. The mean fixes only the product $Na$; two emitters with $a=0.075$ and four emitters with $a=0.0375$ both give $\lambda_{\mathrm{circ}}=0.15$. The within-pulse statistics separate the two factors. Because each emitter contributes at most one photon per pulse, the count is sub-Poissonian; its variance $Na(1-a)$ is lower than its mean $Na$ by
\begin{equation}
 D_{\mathrm{circ}}\equiv\lambda_{\mathrm{circ}}-\mathrm{Var}[K_{\mathrm{circ},\mathrm{tot}}]=Na^2,
 \quad
 N=\frac{\lambda_{\mathrm{circ}}^2}{D_{\mathrm{circ}}}.
 \label{eq:baselineN}
\end{equation}
This deficit reflects the absence of two-photon events from a single emitter and underlies photon-statistics counting \cite{grusmayer_time-resolved_2017,li_en_2022,li_estimation_2023}. In the example, $D_{\mathrm{circ}}=0.01125$ and $0.005625$, and Eq.~\eqref{eq:baselineN} returns $N=2$ and $4$.

With unlimited data, the baseline therefore already determines $N$ in this equal-amplitude model. The question is whether polarization increases the counting information carried by each pulse, and why. The mean signal cannot provide it: replacing every emitter by two copies with the same orientation and half the amplitude leaves the mean count in every channel unchanged at every excitation polarization, since $\mathbb E[K_{pc}]=\sum_i q_{ip}\rho_{ic}$ is linear in the amplitudes. If the two emitters above are perpendicular, $\bm\phi=(0,\pi/2)$, and the four emitters are their duplicates, $\bm\phi=(0,0,\pi/2,\pi/2)$, the two clusters thus have identical polarization-resolved means. Any gain from polarization must come from how it reshapes the within-pulse statistics. 

Excitation polarization reshapes these statistics by modulating the detection probability of each emitter. At linear polarization $p$, the total count is still a sum of independent Bernoulli variables, with mean $\mu_p=\sum_i q_{ip}$, variance $\sigma_p^2=\sum_i q_{ip}(1-q_{ip})$, and
\begin{equation}
 \mu_p-\sigma_p^2=\sum_i q_{ip}^2,\quad
 \frac13\sum_{p=1}^{3}q_{ip}^2=\frac32a_i^2.
 \label{eq:deficit}
\end{equation}
The second relation follows from Eq.~\eqref{eq:excitation} and the equal spacing of the three angles, and holds for every orientation $\phi_i$. Averaged over the three polarizations, the detection probability of each emitter is unchanged [Eq.~\eqref{eq:meanbudget}], but its mean square rises from $a_i^2$ under circular excitation to $3a_i^2/2$. The deficit, being quadratic in the detection probabilities, rises with it. In each polarization, the photons come mainly from the emitters aligned with the excitation field, which sharpens the sub-Poissonian signature. In the example, excitation along $x$ excites only the $x$-oriented emitters: the two-emitter cluster then behaves as a single emitter and never yields two photons, whereas its four-emitter duplicate still can.

For equal amplitudes, the deficit and mean averaged over the three polarizations are
\begin{equation}
 D\equiv\frac13\sum_{p=1}^{3}(\mu_p-\sigma_p^2)=\frac32Na^2,
 \quad \bar\lambda=Na,
 \quad N=\frac{3\bar\lambda^2}{2D},
 \label{eq:countidentity}
\end{equation}
in which the individual orientations cancel. At matched photon budget, $\bar\lambda=\lambda_{\mathrm{circ}}$ and $D=3D_{\mathrm{circ}}/2$; linearly polarized excitation increases the deficit by a factor of $3/2$ for any orientations. It does so without increasing the fluctuations against which the deficit is measured; averaged over isotropic azimuthal angles, the spread of the polarization-dependent means adds exactly the count variance that the larger deficit removes. With a statistical error that decreases as $M^{-1/2}$, the $3/2$ larger contrast between candidate counts should therefore reduce the pulse budget by about $(3/2)^2=2.25$, which Sec.~\ref{sec:inference} evaluates directly.

Emission polarization acts differently. It leaves the number of detected photons unchanged but labels each photon by channel. Under circular excitation with equal amplitudes, the channel-resolved PMF separates into a total-count factor, which depends only on $N$ and $a$, and an allocation factor, which depends only on $N$ and the orientations $\bm\phi=(\phi_1,\ldots,\phi_N)$:
\begin{equation}
 \begin{aligned}
 \Pr(\bm K=\bm k\mid N,a,\bm\phi)
 ={}&\Pr(K_{\mathrm{tot}}=k_{\mathrm{tot}}\mid N,a)\\
 &\times\Pr(\bm K=\bm k\mid K_{\mathrm{tot}}=k_{\mathrm{tot}},N,\bm\phi),
 \end{aligned}
 \label{eq:channelfactorization}
\end{equation}
where $\bm K$ is the channel-resolved readout with the label $p=\mathrm{circ}$ omitted. 

The baseline records only the first factor. The second is independent of $a$. With equal detection probabilities, the probability that a given set of $k_{\mathrm{tot}}$ emitters is detected while the others are not, $a^{k_{\mathrm{tot}}}(1-a)^{N-k_{\mathrm{tot}}}$, is the same for every set; conditioning on $k_{\mathrm{tot}}$ therefore makes all sets equally likely and removes $a$, and the channels the photons enter then depend only on the orientations. Channel-resolved readout therefore adds count information that the total-count statistics do not contain.

Single-photon pulses carry none of this information. With equal detection probabilities, the photon is equally likely to come from any emitter, so it enters channel $c$ with probability $\bar r_c=N^{-1}\sum_i r_c(\phi_i)$. The same $\bar r_c=\mathbb E[K_c]/\mathbb E[K_{\mathrm{tot}}]$ is the fraction of all detected photons recorded in channel $c$, which is measured directly and is unchanged by duplicating the emitters.

Count information first appears in two-photon pulses, whose photons must come from two distinct emitters. We characterize these pulses by the probability that both photons enter the same channel,
\begin{equation}
 P_{\mathrm{same}}\equiv\sum_{c=1}^{4}\Pr(K_c=2\mid K_{\mathrm{tot}}=2).
 \label{eq:psame}
\end{equation}
To see how $P_{\mathrm{same}}$ depends on $N$, consider instead two photons drawn independently from the mean fractions $\bar r_c$; the probability that they enter the same channel is $u\equiv\sum_c\bar r_c^2$. This probability has two contributions. With probability $1-1/N$, the two photons come from distinct emitters, as in a real two-photon pulse; since all pairs of distinct emitters are equally likely, their same-channel probability is $P_{\mathrm{same}}$. With probability $1/N$, both come from the same emitter, a scenario that cannot occur in a real pulse because no emitter contributes two photons; their same-channel probability is then $\kappa\equiv\sum_c r_c(\phi_i)^2=1/4+V^2/8$, which is independent of $\phi_i$ because the channel angles are equally spaced. Therefore
\begin{equation}
 u=\Bigl(1-\frac1N\Bigr)P_{\mathrm{same}}+\frac{\kappa}{N},\quad
 N=\frac{\kappa-P_{\mathrm{same}}}{u-P_{\mathrm{same}}}.
 \label{eq:emissioncountidentity}
\end{equation}
Real two-photon pulses lack the second contribution, and the resulting difference between $P_{\mathrm{same}}$ and $u$ reveals $N$. The identity involves only the measurable quantities $u$ and $P_{\mathrm{same}}$, and the known visibility $V$; neither the orientations nor the amplitude enters.

The identity rests on the gap
\begin{equation}
 u-P_{\mathrm{same}}=\frac{1}{N-1}\sum_{c=1}^{4}\operatorname{Var}_i[r_c(\phi_i)],
 \label{eq:emissioncontrast}
\end{equation}
where $\operatorname{Var}_i$ denotes the variance over the $N$ emitters, normalized by $N$. It grows with the spread of the emitters' orientations and decreases as $1/(N-1)$, because same-emitter draws make up a smaller fraction of all independent draws in larger clusters.

Consider again the two perpendicular emitters, $\bm\phi=(0,\pi/2)$, and their four-emitter duplicate, $\bm\phi=(0,0,\pi/2,\pi/2)$. Photons from the $x$- and $y$-oriented emitters favor the $0^\circ$ and $90^\circ$ channels, respectively, so for $V=0.958$, photons from a perpendicular pair enter the same channel with probability $0.135$, whereas photons from a parallel pair do so with probability $\kappa=0.365$. In the two-emitter cluster, every two-photon pulse comes from the perpendicular pair, giving $P_{\mathrm{same}}=0.135$; in the four-emitter cluster, two of the six pairs are parallel, raising $P_{\mathrm{same}}$ to $0.212$. Both clusters have $\bar r_c=1/4$ and $u=1/4$, and Eq.~\eqref{eq:emissioncountidentity} returns $N=2$ and $4$.

The identity becomes indeterminate if all emitters are parallel or if $V=0$: then $u=\kappa=P_{\mathrm{same}}$, and the allocation carries no count information. Unlike the orientation-independent excitation deficit, this emission signature depends on the configuration: nearly parallel emitters give a small gap and hence an unstable finite-data estimate. In more general cases, with unequal amplitudes, linear excitation, or the tilt-dependent visibility of three-dimensional dipoles, the identity no longer holds exactly; these cases are evaluated numerically in Sec.~\ref{sec:inference}.

\section{Likelihood inference and numerical results}
\label{sec:inference}

\begin{table*}[t]
\caption{Pulse budget $M_{95}$ at which 95\% of the photon records yield the correct count, in thousands of total pulses, for the four protocols of Sec.~\ref{sec:protocols}. Each entry uses 1000 configurations with three photon records each; square brackets give 95\% intervals obtained by resampling the configurations.}
\label{tab:M95}
\begin{ruledtabular}
\begin{tabular}{llcccc}
Model & $N$ & Baseline & 4em & 3ex & 3ex+4em \\
\colrule
 & 2 & 9.03 [8.45, 9.67] & 7.92 [7.20, 8.88] & 4.37 [4.06, 4.69] & 4.19 [3.66, 4.78] \\
Equal amplitudes & 3 & 87.0 [79.1, 93.9] & 67.4 [61.6, 74.9] & 29.2 [26.4, 31.9] & 22.9 [20.5, 24.9] \\
 & 4 & 274 [248, 301] & 234 [211, 253] & 101 [93.1, 110] & 76.8 [68.0, 82.4] \\
 & 5 & 707 [656, 761] & 578 [542, 624] & 269 [253, 285] & 176 [160, 198] \\
\colrule
 & 2 & 10.2 [9.30, 11.1] & 10.8 [9.59, 12.0] & 5.77 [5.24, 6.49] & 4.75 [4.40, 5.15] \\
Unequal amplitudes & 3 & 110 [98.7, 120] & 104 [94.6, 113] & 45.7 [40.4, 52.0] & 40.6 [37.3, 46.0] \\
 & 4 & 466 [421, 513] & 438 [391, 486] & 206 [184, 231] & 169 [153, 184] \\
 & 5 & 1340 [1180, 1530] & 1260 [1140, 1390] & 604 [555, 699] & 522 [472, 568] \\
\colrule
 & 2 & 9.12 [8.59, 9.65] & 8.56 [7.81, 9.41] & 5.24 [4.56, 5.88] & 4.96 [4.51, 5.35] \\
Three-dimensional & 3 & 109 [98.0, 120] & 84.6 [77.5, 92.1] & 45.7 [41.0, 50.1] & 33.8 [29.8, 36.9] \\
 & 4 & 453 [400, 507] & 352 [322, 386] & 206 [181, 231] & 107 [99.0, 115] \\
 & 5 & 1430 [1210, 1650] & 1020 [929, 1130] & 694 [625, 762] & 297 [278, 321] \\
\end{tabular}
\end{ruledtabular}
\end{table*}

We infer $N$ by maximum likelihood. For a given protocol, let $h_p(\bm k)$ be the number of pulses at polarization $p$ with readout $\bm k$, including empty pulses; $\bm k$ has one component for total-only readout and four for channel-resolved readout [Eq.~\eqref{eq:readout}]. For each candidate $N$, the log-likelihood is
\begin{equation}
 \ell(N,\xi_N)=\sum_p\sum_{\bm k}h_p(\bm k)\log P_p(\bm k\mid N,\xi_N),
 \label{eq:loglik}
\end{equation}
where the sum over $p$ runs over the polarizations of the protocol and $P_p$ is the PMF of Eq.~\eqref{eq:jointpmf}. The parameters $\xi_N$ are the amplitudes and, for protocols sensitive to orientation, the dipole orientations; their feasible set $\Xi_N$ encodes the model under test. The estimate is
\begin{equation}
 \hat N=\operatorname*{arg\,max}_{N\in\{1,\ldots,6\}}\ \max_{\xi_N\in\Xi_N}\ell(N,\xi_N).
 \label{eq:Nhat}
\end{equation}
We compare candidates by their maximized likelihoods without a complexity penalty. Such a penalty would be essential if a candidate with more emitters could reproduce every cluster of fewer emitters, for example by switching one emitter off, because it would then always fit at least as well; the amplitude constraints below exclude this. Candidates with more emitters nevertheless remain more flexible, which favors overestimation in short records, and the accuracies reported below include this effect.
We evaluate the PMF of Eq.~\eqref{eq:jointpmf} by adding the emitters one at a time (iterative convolution), which avoids enumerating all combinations of single-emitter outcomes. For each candidate, we then maximize $\ell$ over $\xi_N$ with a gradient-based local optimizer started from at least eight points: orientations estimated from the polarization-dependent mean counts, random points, duplicated solutions of smaller candidates, and the solutions found at the preceding budget and for the corresponding total-only or channel-resolved protocol. The best result is retained.

We study three models, each used both to generate data and to fit them. The first is the equal-amplitude model of Sec.~\ref{sec:theory}: in-plane emitters share an unknown amplitude $a$, and $\xi_N$ contains $a$ and, for orientation-sensitive protocols, the azimuthal angles $\phi_i$. The second allows unequal amplitudes: each in-plane emitter has its own amplitude $a_i$, which replaces $a$ in $\xi_N$, subject to $\max_i a_i/\min_i a_i\leq2$. This bound defines the emitters being counted. Without it, counting would be ill-posed, because an arbitrarily dim emitter changes the photon statistics arbitrarily little and no finite record can reveal it. 
With it, counting remains well posed: for unequal amplitudes, $\bar\lambda=\sum_i a_i$ and $D=3\sum_i a_i^2/2$, so the identity of Eq.~\eqref{eq:countidentity} returns the effective number $(\sum_i a_i)^2/\sum_i a_i^2\leq N$ instead of $N$; the amplitude-ratio bound keeps this number above $N-1$ for all $N\leq8$, and the second moments therefore still separate the candidate counts in principle.
The bound is also a reasonable assumption for experiments. It still allows a twofold spread in brightness, and since the deficit that carries the count information scales as $a_i^2$, an emitter at half the maximum amplitude contributes only a quarter as much to it as the brightest one; dimmer emitters contribute even less and are hard to separate from background. In both planar models, $\Xi_N$ also bounds each amplitude by $a_{\max}=f_{xy}/2=0.175$, half the collection fraction $f_{xy}=0.350$ of an in-plane dipole \cite{axelrod_carbocyanine_1979,Fourkas2001,Axelrod2012,Backer2014}, because the peak detection probability $2a_i$ cannot exceed $f_{xy}$.

The third is a three-dimensional model, in which each dipole has a polar angle $\theta_i$ from the optical axis and an azimuthal angle $\phi_i$, and all emitters share an intrinsic peak excitation--emission probability $A\leq1$. The transverse excitation field acts on the in-plane projection of each dipole, so Eq.~\eqref{eq:excitation} holds with the effective amplitude $a_i=Af_i\sin^2\theta_i/2$, where $f_i=f_z\cos^2\theta_i+f_{xy}\sin^2\theta_i$ is the collection fraction and $f_z=0.227$ is that of an axial dipole \cite{axelrod_carbocyanine_1979,Fourkas2001,Axelrod2012,Backer2014}. Tilt also lowers the polarization visibility in Eq.~\eqref{eq:routing} to $V_i=Vf_{xy}\sin^2\theta_i/f_i$. Total-only data depend on $\theta_i$ and $A$ only through $a_i$, which we fit as in the unequal-amplitude model; channel-resolved fits use $a_i$, $\phi_i$, and a common $A$, from which the tilts follow. Longitudinal polarization components of the excitation field are neglected.

We consider true counts $N=2$--5, the moderate overlaps expected when the emitter density is raised to accelerate SMLM; larger clusters, whose thresholds grow steeply with $N$ (Table~\ref{tab:M95}), are outside our scope. The candidate range in Eq.~\eqref{eq:Nhat} extends one emitter beyond the largest true count, so overestimates of $N=5$ are limited to a single emitter. For each model and each $N$, we draw 1000 independent configurations, i.e., sets of true orientations and amplitudes, and simulate three independent photon records for each.
Azimuthal angles are uniform on $[0,\pi)$. In the unequal-amplitude model, relative amplitudes are drawn log-uniformly between 1 and 2; in the three-dimensional model, orientations are isotropic ($\cos\theta_i$ uniform on $[-1,1]$). All configurations are scaled to $\bar\lambda=0.15$ detected photons per pulse, a value used only to generate data and unknown to the estimator. In the three-dimensional model, $\bar\lambda=\sum_i a_i$ fixes $A=2\bar\lambda/\sum_i f_i\sin^2\theta_i$, and a configuration is retained only if $A\leq1$ and its effective amplitudes differ by at most a factor of 2.
Each photon record is a single simulated pulse sequence, and the record for a budget $M$ consists of its first $M$ pulses, so shorter records are contained in longer ones, as in an experiment that keeps acquiring. Total-only records are obtained by summing the channel counts of the same simulated pulses, so total-only and channel-resolved protocols are compared on identical photons, and all four protocols share the configurations and budgets.

We evaluate the accuracy, i.e., the fraction of records with $\hat N=N$, on a geometric grid of 17 budgets for each $N$, spaced by factors of about 1.25 and spanning the thresholds of all four protocols.
Because the measured accuracy can dip by chance between neighboring budgets, we replace it with the closest nondecreasing sequence and define the threshold $M_{95}$ as the budget, interpolated linearly in $\log M$, at which this accuracy reaches 0.95. To estimate the uncertainty arising from the finite set of emitter configurations, we draw 1000 configurations at random from the original 1000, allowing repeats, keep the three records of each, and recompute $M_{95}$; repeating this 10,000 times gives the 95\% intervals in Table~\ref{tab:M95} as the 2.5th and 97.5th percentiles. Each resample uses the same emitter configurations for all four protocols, so configurations that are hard for every protocol raise all thresholds together and largely cancel in their ratios.

Table~\ref{tab:M95} lists the thresholds. They are stable with respect to the ensemble size: increasing it from 750 to 1000 configurations changed the thresholds by 2.3\% on average, compared with a mean half-width of 9.7\% for the 95\% intervals.
The thresholds rise by factors of $\sim$40--160 from $N=2$ to $N=5$, because the deficits of neighboring counts approach each other as $N$ grows. Because the three models differ in both their ensembles and their fitted parameters, we compare protocols within each model rather than thresholds across models.

\begin{figure}[t!]
\includegraphics{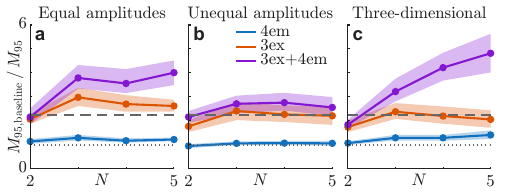}
\caption{Pulse-budget speedup over the circular-excitation baseline, $M_{95}^{\,\mathrm{baseline}}/M_{95}$, for channel-resolved readout (4em), three-angle excitation (3ex), and their combination (3ex+4em). (a) Equal amplitudes, (b) unequal amplitudes, and (c) three-dimensional orientations. Shaded areas are 95\% intervals obtained by resampling the 1000 configurations. The dashed line marks the $(3/2)^2=2.25$ reduction predicted in Sec.~\ref{sec:theory}; the dotted line marks no gain.}
\label{fig:speedup}
\end{figure}

Figure~\ref{fig:speedup} shows the speedups relative to the baseline. Three-angle excitation reduces the threshold by factors of 1.74--2.98 across all twelve groups, with a mean of 2.28, in agreement with the $(3/2)^2=2.25$ predicted in Sec.~\ref{sec:theory}. Although that estimate assumed equal in-plane amplitudes, unequal amplitudes and tilted dipoles give similar factors, with means of 2.17 and 2.10, compared with 2.59 for equal amplitudes.

Emission labels act differently. Alone, they speed up counting by factors of only 1.07--1.41 for equal amplitudes and in three dimensions, and not at all in the unequal-amplitude model (0.95--1.07). This weak standalone gain is consistent with Sec.~\ref{sec:theory}: the allocation signal is carried by multiphoton pulses alone and decreases as $1/(N-1)$, and its count identity requires equal pair probabilities, which unequal amplitudes break. In the unequal-amplitude model, using the channel counts also requires fitting $N$ azimuthal angles in addition to the $N$ amplitudes. This extra flexibility favors candidates with more emitters while adding little count information; at $N=2$, 4em is no faster than the baseline, with a threshold ratio of 0.949 [0.865, 1.03].

Combined with three-angle excitation, emission labels help more, and increasingly so with $N$ for equal amplitudes and tilted dipoles.
Relative to 3ex, 3ex+4em reduces the threshold by factors rising from 1.04 at $N=2$ to 1.53 at $N=5$ for equal amplitudes (Fig. ~\ref{fig:speedup}a), by 1.13--1.22 without a clear trend for unequal amplitudes (Fig. ~\ref{fig:speedup}b), and by factors rising from 1.06 to 2.34 in three dimensions (Fig. ~\ref{fig:speedup}c), giving overall speedups over the baseline of up to 4.0, 2.8, and 4.8, respectively.
This growth may arise because, with 3ex alone, the orientations must be inferred from the same counts that determine $N$; the channel labels supply independent orientation information, leaving more of the excitation statistics for counting, and their value grows with the number of orientations to be determined.

\section{Discussion and Conclusion}

Our results show that polarization reduces the pulse budget for counting overlapping dipole emitters by reshaping their within-pulse photon statistics, since mean polarization-resolved signals cannot count. It does so in two complementary ways: linearly polarized excitation strengthens the sub-Poissonian signature of the count for any emitter orientations, and polarization-resolved detection reveals, through the channels in which the photons of multiphoton pulses are recorded, how many distinct emitters share the light. Combined, they reduce the pulse budget by factors of $\sim2$ to 5.

The required optics are already used for single-molecule orientation measurements~\cite{Backer2016,Rimoli2022,Beausang2013}. The achievable speed is set by the detector. Scanning photon-number-resolving modules~\cite{achilles_fiber-assisted_2003} at megahertz repetition rates would acquire the budgets of Table~\ref{tab:M95} in milliseconds to about a second per spot. Current widefield detectors are limited by frame rate in qCMOS sensors, because each exposure must contain a single pulse, and by detection efficiency in gated SPAD arrays~\cite{ulku_512_2019}. 
With qCMOS sensors, a practical scheme first locates regions with overlapping emitters in a full-frame image and then counts within them using a narrow subarray, whose frame rate depends only on the number of rows read. 
For instance, with the four channels side by side in a 32-row subarray, read at the specified 6610 frames/s in the standard scan of a commercial qCMOS camera (ORCA-Quest 2, Hamamatsu), 3ex+4em counts two or three emitters in 0.6--6~s, compared with 1.4--17~s for the baseline.
These times, during which each emitter yields a few hundred to about two thousand detected photons, are already comparable to those of SMLM with long exposures~\cite{dai_optical_2016,diekmann_optimizing_2020}. Because a single pulse per frame also caps the photon rate of each emitter, matching fast SMLM will require future widefield detectors that resolve photon numbers at megahertz rates, with which 3ex+4em would count three emitters in 23--41~ms, about the length of a single exposure.

A future direction of this work is to extend the analysis to the more complex molecular configurations encountered in biological imaging. Our estimates assume fixed dipoles, ideal readout, and no background or blinking. Rotational mobility~\cite{zhang_single-molecule_2024,brasselet_single-molecule_2025} weakens both mechanisms by reducing the excitation modulation and the emission visibility. Because the mobility differs among emitters, it also adds fitted parameters that, like unequal amplitudes, favor overcounting; polarization-resolved detection, however, also measures these parameters independently of counting~\cite{Rimoli2022,Beausang2013} and may then help more. The statistics are further distorted by slow orientation changes, which make them drift over time; by fast orientation changes within the fluorescence lifetime, which decouple the emission polarization from the excitation polarization; and by blinking, bleaching, background, and energy transfer between closely spaced emitters. Orientation changes faster than the pulse period, including those within the fluorescence lifetime, and background with a known rate can be included in the forward model without changing the inference, whereas slower orientation changes, blinking, bleaching, and energy transfer break the assumption of independent, stationary emitters on which the model rests; quantifying all of these effects will require experiments on samples with known emitter numbers, such as DNA origami.

Because polarization-encoded counting requires only polarization optics and photon-number-resolved detection and does not otherwise constrain the imaging, it can be combined with existing single-molecule techniques and optimized alongside them.
The framework developed here can be used to choose the excitation states, their pulse allocation, and the emission channels that minimize the pulse budget.
The polarization channels could also be complemented by other emitter-dependent labels, such as the image-plane pixels used in antibunching-based imaging~\cite{schwartz_superresolution_2013,tenne_super-resolution_2019,israel_quantum_2017} or engineered dipole-spread functions~\cite{Zhang2022,chen_resolving_2025}. Counting also supplies the emitter number that analyses of resolution take as known~\cite{Tsang2016,dingilian_quantifying_2026}, and the same polarization-resolved photons carry information on orientation and position, so that an unresolved spot could be counted, oriented, and localized jointly. How closely such joint measurements can approach the quantum limits of counting~\cite{lu_quantum-optimal_2018,zanforlin_optical_2022} and of localization and orientation estimation~\cite{Backlund2018,Zhang2020,Zhang2021} remains open.

In conclusion, polarization, conventionally exploited to measure the orientations of dipole emitters, also carries information on their number. By reshaping the photon statistics of each pulse, it reduces the pulse budget for counting overlapping emitters severalfold with existing optics, and it opens measurement design as a route to faster counting.

\begin{acknowledgments}
This work was supported by start-up funds from Texas A\&M University.
The author used AI tools (Codex and Claude) to assist in optimizing the implementation of the estimator, substantially reducing its computational runtime. The resulting code was reviewed and checked for accuracy and consistency with the model described in this manuscript.
\end{acknowledgments}

% \section*{Data availability} The original MATLAB implementation and the optimized Python implementation of the estimator are publicly available in Ref. \hl{[cite]}.

% \appendix

\bibliography{references}% Produces the bibliography via BibTeX.

\end{document}